\documentclass[aip,preprint]{revtex4-1}

\usepackage{graphicx}
\usepackage[english]{babel}
\usepackage[T1]{fontenc}
\usepackage[colorlinks,linkcolor=blue,citecolor=blue,urlcolor=blue]{hyperref}
\usepackage{amsmath,amssymb}

\begin{document}

\title{Importance of accurate consideration of the electron inertia 
in hybrid-kinetic simulations of collisionless plasma turbulence: 1. The 2D limit}
\author{Neeraj Jain}
\email[]{neeraj.jain@tu-berlin.de}
\affiliation{Center for Astronomy and Astrophysics, Technische Universit\"at Berlin, 10623 Berlin, Germany}
\author{Patricio Mu\~noz}
\affiliation{Max Planck Institute for Solar System Research, 37077 G\"ottingen, Germany}
\author{Meisam Farzalipour Tabriz}
\author{Markus Rampp}
\affiliation{Max Planck Computing and Data Facility, 85748 Garching, Germany}
\author{J\"org B\"uchner}
\affiliation{Max Planck Institute for Solar System Research, 37077 G\"ottingen, Germany}
\affiliation{Center for Astronomy and Astrophysics, Technische Universit\"at Berlin, 10623 Berlin, Germany}

\date{\today}

\begin{abstract}
The dissipation mechanism of the magnetic energy in turbulent collisionless space 
and astrophysical plasmas is still not well understood. 
Its investigation requires efficient kinetic simulations of the energy transfer in collisionless plasma turbulence.
In this respect, hybrid-kinetic simulations, in which ions are treated as particles and electrons as an inertial fluid, have begun to attract a significant interest recently.
Hybrid-kinetic models describe both ion- and electron scale processes by ignoring 
electron kinetic effects so that they are computationally much less demanding compared to fully kinetic plasma models. 
Hybrid-kinetic codes solve either the Vlasov equation for the ions (Eulerian 
Vlasov-hybrid codes) or the equations of motion of the ions as
macro-particles (Lagrangian Particle-in-Cell (PIC)-hybrid codes).
They consider the inertia of the electron fluid using different 
approximations. 
We check the validity of these approximations by employing our recently 
massively parallelized three-dimensional PIC-hybrid code CHIEF
which considers the electron inertia without any of the common approximations.
In particular we report the results of simulations of two-dimensional 
collisionless plasma turbulence. 
We conclude that the simulation results obtained using hybrid-kinetic 
codes which use approximations to describe the electron inertia 
need to be interpreted with caution. 
We also discuss the parallel scalability of CHIEF, to the best of our knowledge,
the first PIC-hybrid code which without approximations describes the
inertial electron fluid.

\end{abstract}


\maketitle 

\section{\label{sec:intro}Introduction}

The nature of collisionless plasma turbulence, from large (fluid) down to small (kinetic) scales, is a key unsolved problem in plasma physics \cite{kiyani2015}. It finds its importance in many physical situations ranging from hot laboratory to astrophysical plasmas \cite{Schekochihin2007,fujisawa2021}. In these predominantly collisionless plasmas, turbulence develops from fluid to kinetic scales where its energy is finally dissipated into heat. 
Thus, the understanding of heating and dissipation in collisionless plasma 
requires kinetic studies of the plasma turbulence. 
This is particularly true for the study of the turbulent solar wind 
plasma~\cite{marsch2006} which 
provides an excellent natural laboratory for turbulence studies in a broad 
range of parameters~\cite{bruno2013}. 
In fact, recent {\it in situ} spacecraft measurements in the solar wind 
have provided a wealth of data which extended the observational 
basis to kinetic scales~\cite{alexandrova2013,chen2016,goldstein2015,matteini2020}. 
Recent observations in the inner solar corona by the Parker Solar Probe 
have revealed new kinetic scale features of the collisionless turbulence of the solar wind plasma~\cite{duan2021,kasper2021,bowen2020}. 
On the other hand, numerical simulations of collisionless kinetic plasma turbulence 
phenomena have helped to better understand step-by-step the physical
processes behind it~\cite{matteini2020,cerri2019,groselj2017,jain2021,franci2018}. 

Plasma kinetic scales range from ion- down to electron scales. 
An ideal numerical simulation of the kinetic plasma turbulence would,
therefore, have to cover the full range of plasma kinetic scales, from 
sufficiently above ion scales to below the electron scales,
treating both ions and electrons kinetically. 
Fully kinetic particle-in-cell (PIC) codes, which treat ions and electrons as particles, as well as Vlasov-codes, which solve for the evolution
of phase-space distribution functions of electrons and ions, are, however, computationally 
very demanding. 
They are typically carried out for artificially modified physical parameters, 
like reduced ion-to-electron mass ratios and for small spatial domains.
Recently three-dimensional PIC-code simulations with more and
more realistic physical parameters are becoming feasible with the 
increasing computational power.
Nevertheless, the inherent noise associated with the always reduced 
number of macro-particles in PIC-code simulations makes the particle 
statistics at the smallest scales unreliable. 
The alternative, a direct numerical solution of the Vlasov equations for the
ion- and electron distribution functions, is very expensive in terms of 
computer memory for an appropriate velocity space 
resolution~\cite{buchner2005,kormann2019}. 

One way out of this situation is to understand the processes in parts by 
simulating certain aspects of physics while neglecting others.
This implies the use of a variety of approximations utilized by
different simulation models. 
A comprehensive understanding is then obtained by combining various aspects of
these simulations. 
Even if we were able to simulate all the scales simultaneously, a physicist's 
approach would be to understand the physical processes in parts and then 
develop an overall understanding by comprehending the essential 
physics.

Hybrid-kinetic plasma simulation models treat electrons as a fluid but ions 
kinetically. This relaxes some of the computational constraints of the fully 
kinetic simulation models and thus reduces the computational 
cost~\cite{Lipatov2002}. 
Hybrid-kinetic models do not describe the kinetic effects associated with 
electrons. 
Such models are suitable to study processes in which ion kinetic effects 
are more important than electron kinetic effects. 
The majority of hybrid-kinetic simulation codes further simplify 
their algorithms by assuming an inertia-less electron fluid. 
This allows a direct calculation of the electric field from the electron's 
momentum equation without the need for solving a differential equation. 
On the other hand, this limits their validity to scales 
much larger than the electron scales.
Space observations of collisionless plasma turbulence by multi-spacecraft missions (Cluster and MMS) in the Earth's magnetosphere and the solar wind, however, 
revealed electron- and ion-scale breaks in the power spectra
as well as the formation of current sheet (CS) structures from ion to electron scales ~\cite{khabarova2021,wang2018,zhou2019}.  
On the other hand, even hybrid-kinetic simulations with inertia-less 
electrons have already shown that in evolving turbulent CSs, 
the electrons become the main current carriers~\cite{jain2021} and 
that these CSs thin down to the grid scale~\cite{azizabadi2021}. 
This finding implies that the electron mass has to be incorporated 
in hybrid-kinetic plasma models in order to understand the role of electron-scale CSs in collisionless
turbulent plasmas.

Hybrid-kinetic codes with electron inertia have been under development 
over the last six decades. 
First, PIC-hybrid codes considering ions as particles and an
inertial electron fluid were developed. 
They were applied to simulations of collisionless shock waves and magnetic reconnection~\cite{Forslund1971,Hewett1978,Swift1996,Shay1998,Kuznetsova1998,Swift2001,Lipatov2002,Amano2014}. 
Later, Eulerian Vlasov-hybrid codes with an inertial electron fluid were also developed. They solve the Vlasov equation for the ions in the phase 
space~\cite{Valentini2007,Cheng2013}. 
In these previous PIC-hybrid and Vlasov-hybrid codes, the electron inertia is 
considered in the momentum equation of the electron fluid (the generalized 
Ohm's law) under varying degrees of approximations, which are, however, 
not necessarily valid at the electron scales (see Ref. \citet{Munoz2018} for a 
detailed discussion of these approximations and their limits). 
Our recently developed PIC-hybrid code CHIEF 
(Code Hybrid with Inertial Electron Fluid), by contrast, considers the 
electron inertia without any of those approximations~\cite{Munoz2018}. 
In Ref. \citet{Munoz2018} the numerical algorithms of the CHIEF code 
are described as well as the results of the validation of the code against a number 
of physical benchmark problems.
 
We have now updated the code by massively parallelizing it for high performance computing (HPC) on supercomputers. This enables two- and three-dimensional simulations of the
evolution of turbulent plasmas with an electron scale spatial resolution and on 
physically relevant spatial- and temporal scales.

In this paper we report our assessment of the validity of the different 
approximations used to describe the effects of the electron inertia by utilizing the simulation
results obtained by the accurate description of the inertial electron 
fluid in the CHIEF code. In particular, we have simulated the evolution of two dimensional 
collisionless plasma turbulence 
utilizing CHIEF, which is genuinely a three-dimensional code, 
in a quasi-two dimensional mode by reducing the spatial resolution in 
the third dimension to four-grid points.
CHIEF simulations allowing 
variations in all three spatial dimensions will be presented in a forthcoming 
publication.
CHIEF is suitable to assess the validity of describing the electron 
inertia by different approximations as it does not use any of them itself. 

We also compare the results of the CHIEF simulations obtained for inertial and inertia-less electron fluids to understand the role of the electron inertia.
Our simulations have shown that the commonly used approximate 
descriptions of electron inertia are inapplicable to simulate the
physical processes of collisionless plasma turbulence in two dimensions. 
The paper is organized as follows:
in Sec.~\ref{sec:model} we present a short description of the hybrid-kinetic plasma model 
used by CHIEF. 
 In Sec.~ \ref{sec:numerical_implementation} we discuss the numerical 
implementation of the MPI-based parallelization of CHIEF. 
The simulation setup is presented in Sec.~\ref{sec:setup}. 
Sec.~\ref{sec:simulations} contains the results of simulations of collisionless plasma turbulence and an assessment of the validity of 
the commonly used electron-inertia related approximations.
We summarize our results in Sec. \ref{sec:conclusion}. 
In Appendix~\ref{sec:performance} we present a basic performance 
assessment of CHIEF and of its scalability using a few thousand 
CPU cores for spatial grid-resolutions up to 4096$\times$4096$\times$4 
grid points and for up to 500 protons per cell. 
The discretization of a second order elliptic partial differential equations 
without cross derivative terms used by CHIEF is presented in 
Appendix~\ref{sec:discretizaion_elliptic}.

\section{Hybrid-kinetic plasma model with electron inertia\label{sec:model}}

The PIC-hybrid code CHIEF treats ions as Lagrangian macro-particles modelled via the Particle-in-Cell (PIC) method and electrons as an Eulerian fluid with finite mass and temperature. The position $\vec{x}_i^{\;p}$ and velocity $\vec{v}_i^{\;p}$ of each ion macro-particle (the index ``$p$'' denotes each numerical macro-particle, while ``$i$'' indicates ions as a plasma species), representing a large number of physical ions, are obtained from:

\begin{align}
	\frac{d\vec{x}_i^{\;p}}{dt}    & =\vec{v}_i^{\;p}, \label{eq:xp}                                \\
	m_i\frac{d\vec{v}_i^{\;p}}{dt} & =e(\vec{E}^{\;p} + \vec{v}_i^{\;p}\times \vec{B}^{\;p}). \label{eq:vp}
\end{align}
Here, $e$ and $m_i$ are the fundamental charge and mass of a physical ion, respectively. The electric and magnetic field at the particles' position, $\vec{E}^{\;p}$ and $\vec{B}^{\;p}$ respectively, are calculated from the electric and magnetic fields on the numerical grid employing an appropriate weighting scheme. The ion number density $n_i$ and the ion current density $\vec{\jmath}_i$ on the numerical grid is obtained from the particles' positions and velocities employing the same weighting scheme (See \citet{Munoz2018} for details). The electric and magnetic fields on the grid, $\vec{E}$ and $\vec{B}$, are obtained using $n_i$,  $\vec{\jmath}_i$ and the ion bulk velocity $\vec{u}_i=\vec{\jmath}_i/en_i$.
The electric field on the grid, $\vec{E}$, is calculated from the generalized Ohm's law with the electron inertial and resistive terms included.
\begin{align} \label{eq:e_ohm}
	\vec{E} & = -\vec{u}_e\times \vec{B} -\frac{1}{en}\nabla p_e - \frac{m_e}{e}\left(\frac{\partial \vec{u}_e}{\partial t}+(\vec{u}_e\cdot\vec{\nabla})\vec{u}_e\right)  + \eta \vec{\jmath}.
\end{align}
Here, $\vec{u}_e=\vec{u}_i-\vec{j}/e n$ is the electron bulk velocity, $p_e$ the scalar electron pressure, $n=n_i$ the electron and ion number density, $\vec{\jmath}=\nabla\times\vec{B}/\mu_0$ the total current density (neglecting the displacement current in the Amp\`{e}re's law, an appropriate assumption for frequencies much lower than the electron plasma frequency) and $\eta=m_e\nu/(e^2n) $ the collisional resistivity with $\nu$ the electron-ion collision frequency and $m_e$ the electron mass.
The magnetic field on the grid is obtained by first solving an evolution equation for the generalized vorticity $\vec{W}=\nabla\times\vec{u}_e-e\vec{B}/m_e$, 

\begin{align} \label{eq:curl_emom}
	\frac{\partial \overrightarrow{W}}{\partial t}
	  & = \vec{\nabla}\times\left [\vec{u}_e\times \overrightarrow{W}\right]-\vec{\nabla}\times\left(\frac{\vec{\nabla} p_e}{m_en}\right)- \vec{\nabla}\times\left(\frac{\nu}{en}\vec{\jmath}\right).
\end{align}
Equation \eqref{eq:curl_emom} is obtained by taking the curl of Eq. \eqref{eq:e_ohm} and using Faraday's law to eliminate the electric field. The magnetic field is then calculated by solving an elliptic partial differential equation (PDE) which is obtained by substituting for $\vec{u}_e$ from Ampere's law $\vec{u}_e=\vec{u}_i-\nabla\times\vec{B}/(\mu_0 e n)$ in the expression for $\vec{W}=\nabla\times\vec{u}_e-e\vec{B}/m_e$.

\begin{align}\label{eq:elliptic_b}
	\frac{1}{\mu_0e}\vec{\nabla}\times\left(\frac{\vec{\nabla}\times\vec{B}}{n}\right)+\frac{e\vec{B}}{m_e} & = \vec{\nabla}\times\vec{u}_i-\overrightarrow{W}. 
\end{align}

Here $\mu_0$ is the magnetic permeability of vacuum. The scalar electron pressure $p_e$ is given by the isothermal equation of state,
\begin{align} \label{eq:eos}
	p_e = n_ek_BT_e,
\end{align}
with $k_B$ the Boltzmann constant and $T_e$ the electron temperature, constant in time. Note that if the temperature $T_e$ 
is spatially uniform, the second term on the RHS of Eq.~\eqref{eq:curl_emom} vanishes. In our hybrid simulation model, we solve Eqs.~\eqref{eq:xp}-\eqref{eq:eos} without making any of the bulk electron inertia related approximations used in other hybrid-kinetic codes \cite{Munoz2018}. 

\section{Numerical implementation and parallelization \label{sec:numerical_implementation}}
The simulation model discussed in Sec.~\ref{sec:model} was numerically implemented 
in the CHIEF code~\cite{Munoz2018} by combining the particle-in-cell (PIC) code ACRONYM~\cite{Kilian2012}
(\url{http://plasma.nerd2nerd.org/})
with an electron-magnetohydrodynamics (EMHD) code~\cite{Jain2003,Jain2006}. 
The numerical algorithms and their implementation and validation were described in detail in Ref.~\cite{Munoz2018}. The CHIEF code now comprises the following updates: 

The code is now fully parallelized by a two-dimensional ("pencil") domain decomposition of the computational grid based on the Message Passing Interface (MPI) standard, and thus overcomes a substantial performance bottleneck of the original version where the EMHD part of the code was still computed in serial on a single processor core. This entailed a replacement of the MUDPACK library of PDE solvers, which we used to solve the elliptic equation (Eq. \eqref{eq:elliptic_b}), by the HYPRE \cite{falgout2002} library which provides a comprehensive collection of general-purpose, scalable linear solvers and multi-grid methods (for details the reader is referred to the HYPRE user manual available at \url{https://hypre.readthedocs.io/en/latest/ch-intro.html}). 
To this end, Eq. \eqref{eq:elliptic_b} was discretized to obtain a system of algebraic equations with periodic boundary conditions (cf. Appendix~\ref{sec:discretizaion_elliptic}). The discretized equations are passed to HYPRE through its linear-algebraic (IJ matrix form) interface which provides access to various sparse-matrix solvers, and preconditioners such as BoomerAMG, AMS, Euclid, GMRES, BiCGSTAB, etc. 
CHIEF provides an interface to the BoomerAMG solver, and a hybrid solver (BoomerAMG preconditioner with PCG, GMRES, or BiCGSTAB solver), and other solvers supported by the HYPRE library can easily be integrated. 

Moreover, we replaced the method for solving the three dimensional partial differential equation, Eq. \eqref{eq:curl_emom}, which in the original CHIEF code employed the one-dimensional flux-corrected transport solver (LCPFCT library \cite{lcpfct1d}) via dimensional splitting, by a more accurate, three dimensional flux-corrected transport solver, LCPFCT3 \cite{lcpfct3d}, which does not rely on the dimensional splitting approximation and in addition turned out to be more scalable for our purposes. We modernized the LCPFCT3 code and added support for the runtime definition of grid size and other parameters, and for the initialization and finalization of time-independent quantities in order to avoid redundant computations in consecutive calls to the LCPFCT3 solver.

\section{Simulation setup \label{sec:setup}}

 We initialize the 2-D simulations in a y-z plane with random phased fluctuations of magnetic field and plasma velocity imposed on an isotropic background plasma of uniform density $n_0$. A uniform magnetic field $B_0\hat{x}$ is applied perpendicular to the simulation plane. Magnetic field fluctuations are calculated from the magnetic vector potential
\begin{eqnarray}
  \tilde{\mathbf{A}}&=&\hat{x}\sum_{k_y,k_z}\delta A_x(k_y,k_z) \sin(k_yy+k_zz+\phi(k_y,k_z)), 
  \end{eqnarray}
where $k_y$ and $k_z$ are wave numbers in the y- and z-direction, respectively, and $\phi$ is the wave-number dependent random phase. The amplitude $\delta A_x$ of the magnetic vector potential is so chosen that the amplitude of the perpendicular magnetic field fluctuation $\delta B_{\perp}=\delta A_x k_{\perp}$ is independent of the wave number $k_{\perp}=(k_y^2+k_z^2)^{1/2}$, i.e., all initialized modes have the same energy. Here we used the subscripts `` $\perp$ '' and `` $||$ '' to denote the components of a vector quantity perpendicular and parallel to the applied magnetic field $B_0\hat{x}$, respectively. Plasma velocity fluctuations have the same form as the magnetic field fluctuations except the random phases so that the magnetic and velocity fluctuations have vanishing correlation but an equi-partition of energy.

We initialize fluctuations in the wave number range $|k_{y,z}d_i| < 0.2$ ($k_{y,z}\neq 0$) to have a root-mean-square value $B_{rms}/B_0=0.24$. Here $d_i=v_{Ai}/\omega_{ci}$, $v_{Ai}=B_0/\sqrt{\mu_0n_0m_i}$ and $\omega_{ci}=eB_0/m_i$ are the inertial length, Alfv\'en velocity and cyclotron frequency of ions, respectively. 
The electron and ion plasma beta are $\beta_e=2\mu_0n_0k_B T_e/B_0^2=0.5$ and $\beta_i=2\mu_0n_0k_BT_i/B_0^2=0.5$, respectively. Here $T_e$ and $T_i$ are electron and ion temperatures, respectively. 
Simulations are carried out for the ion to electron mass ratio $m_i/m_e$= 25. The simulation box size in the y-z plane $102.4 d_i \times 102.4 d_i$ is resolved by $1024 \times 1024$ grid points (giving a grid spacing of 0.5 $d_e$ in each direction where the electron inertial length $d_e=(m_i/m_e)^{-1/2} d_i$) and the number of particles per cell is 500. The time step is $\Delta t$=0.25 $\omega_{ce}^{-1}$. We take the plasma resistivity to be zero. We also carry out simulations for inertia-less electrons to compare with the results obtained for inertial electrons. The grid resolution for the inertia less simulations is 0.1 $d_i$ with 1024 grid points in each of the y- and z-directions. The time step is $\Delta t=0.002 \omega_{ci}^{-1}$ to
satisfy the Courant (CFL) conditions for grid scale whistler waves.
The number of particles per cell is also 500. 
Boundary conditions are periodic in all directions. 

\section{Simulations of collisionless plasma turbulence\label{sec:simulations}}

\begin{figure}[ht!]
\includegraphics[width=0.99\textwidth]{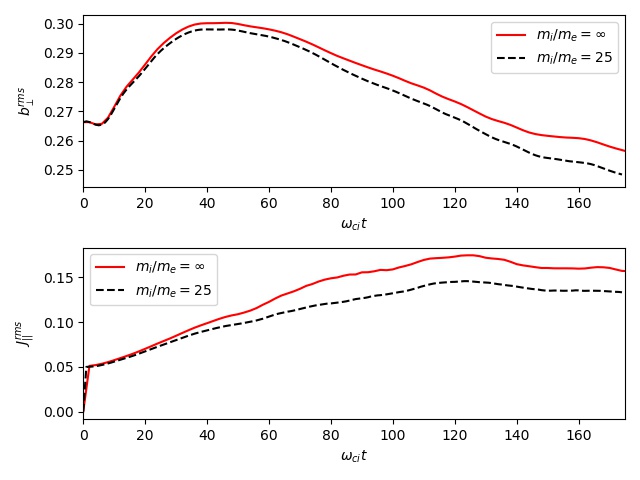}
\caption{Evolution of the rms values of the perpendicular magnetic field ($b_{\perp}/B_0$) and parallel current density ($J_{||}/n_0ev_{Ai}$) for $m_i/m_e=25$ and $m_i/m_e=\infty$ (inertia-less electrons) \label{fig:rms_miDme}}
\end{figure}

Figure \ref{fig:rms_miDme} shows the evolution of the root-mean-square (RMS) values of the perpendicular magnetic field ($b_{\perp}^{rms}$) and parallel current density ($J_{||}^{rms}$) for the cases with inertia-less and inertial ($m_i/m_e=25$) electrons. The initial evolution up to $\omega_{ci}t \approx 10$ is more or less similar for the simulations with inertia-less and inertial electrons. Later, the evolution for the cases with inertia-less and inertial electrons deviate from each other with the latter reaching slightly smaller values. The peak values of $b_{\perp}^{rms}$ and $J_{||}^{rms}$ are reached at around $\omega_{ci}t\approx$40 and 120, respectively, for both the cases of inertial and inertia-less electrons. 

\subsection{Effect of electron inertia on the formation of current sheets}

\begin{figure}[ht!]
	\begin{minipage}[b]{0.49\textwidth}
		\includegraphics[width=0.99\textwidth]{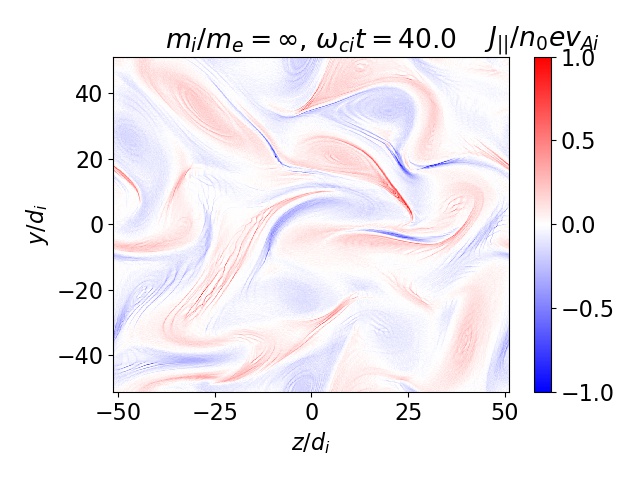}
	\end{minipage}
	\begin{minipage}[b]{0.49\textwidth}
		\includegraphics[width=0.99\textwidth]{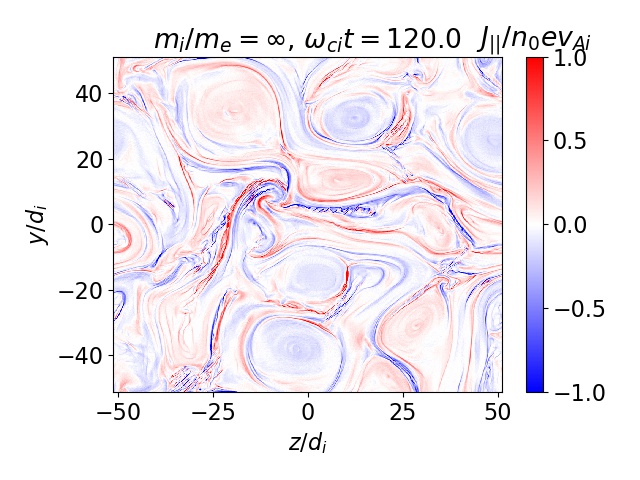}
	\end{minipage}
	\begin{minipage}[b]{0.49\textwidth}
		\includegraphics[width=0.99\textwidth]{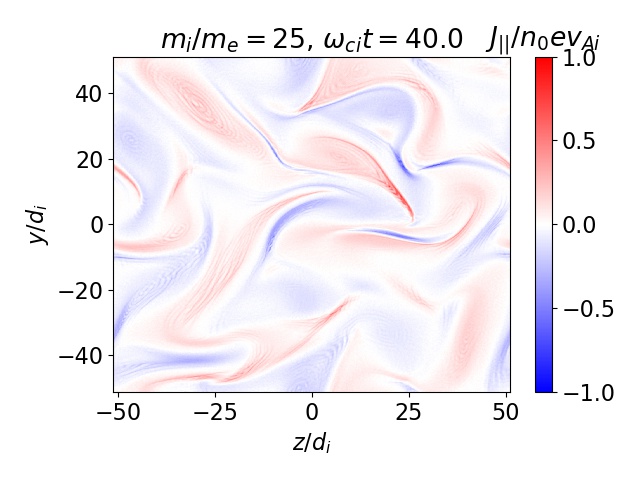}
	\end{minipage}
	\begin{minipage}[b]{0.49\textwidth}
		\includegraphics[width=0.99\textwidth]{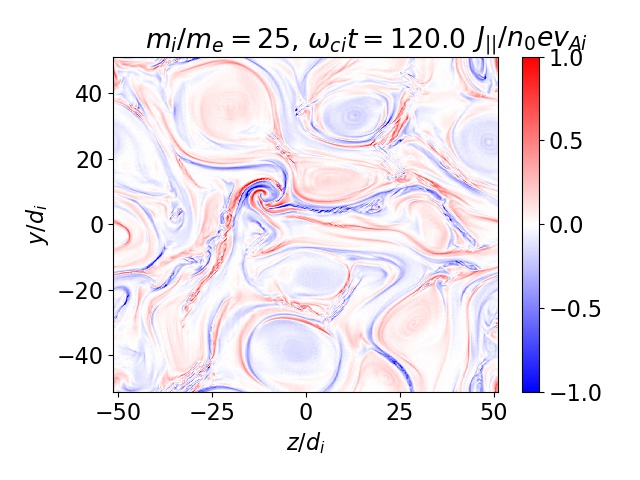}
	\end{minipage}

        \caption{Out-of-plane current density for inertia-less electrons (top row) and $m_i/m_e=25$ (bottom row) at $\omega_{ci}t=40$ (left column) and 120 (right column). \label{fig:jpar_miDme}}
\end{figure}

Fig. \ref{fig:jpar_miDme} shows the current density parallel to the applied magnetic field for the cases of inertial and inertia-less electrons at $\omega_{ci}t$=40 and 120 when $b_{\perp}^{rms}$ and $J_{||}^{rms}$ reach their respective peaks. By $\omega_{ci}t=40$, the power in the initial random Alfv\'enic fluctuations in the wave number range $k_{y,z}d_i < 0.2$ has cascaded down to much shorter scales to form current sheets in the configuration space. The current sheets evolve further and show development of turbulence within them at $\omega_{ci}t$=120 possibly due to the growth of current sheet instabilities.

\begin{figure}[ht!]
	\begin{minipage}[b]{0.49\textwidth}
		\includegraphics[width=0.99\textwidth]{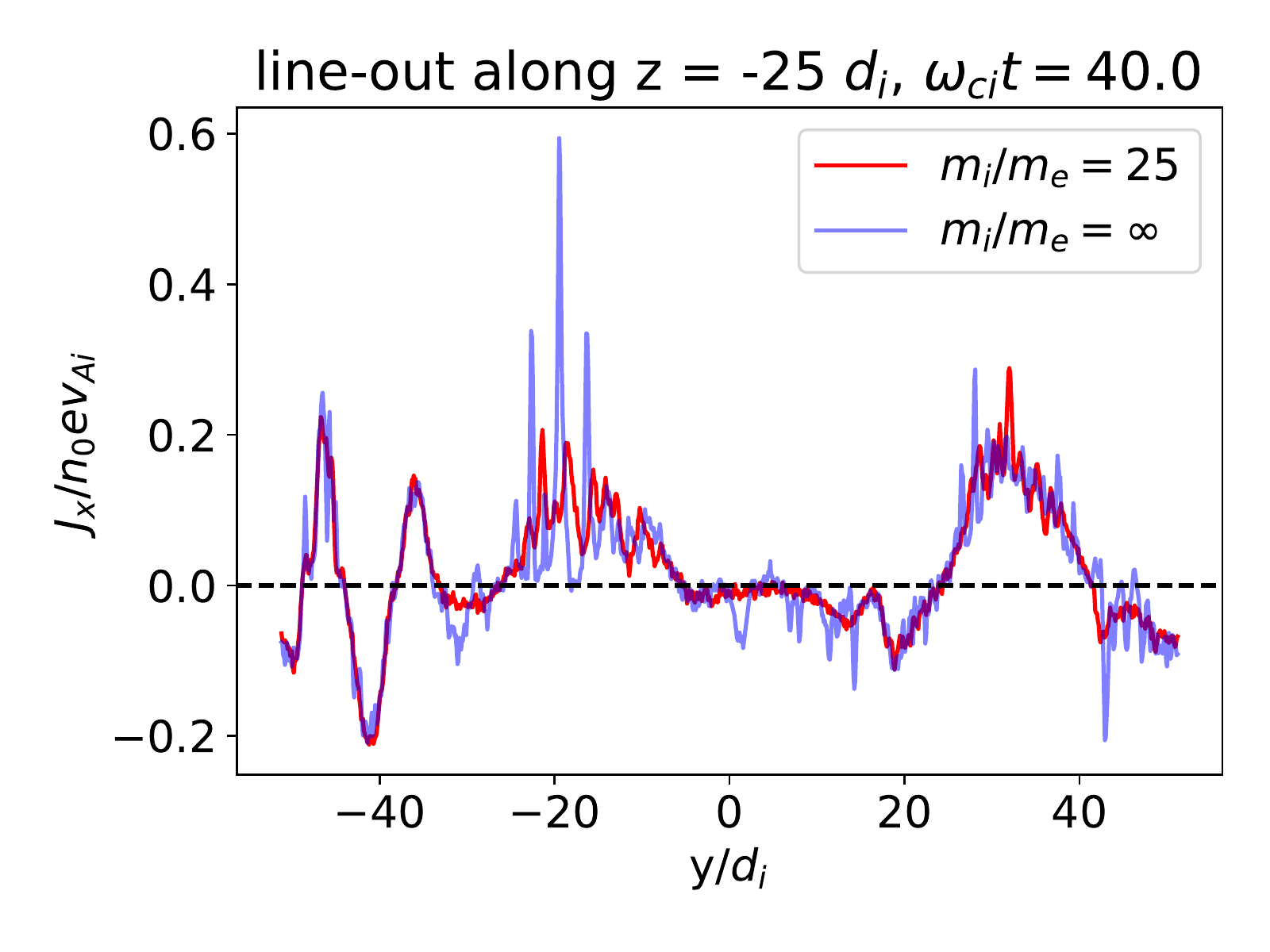}
	\end{minipage}
	\begin{minipage}[b]{0.49\textwidth}
		\includegraphics[width=0.99\textwidth]{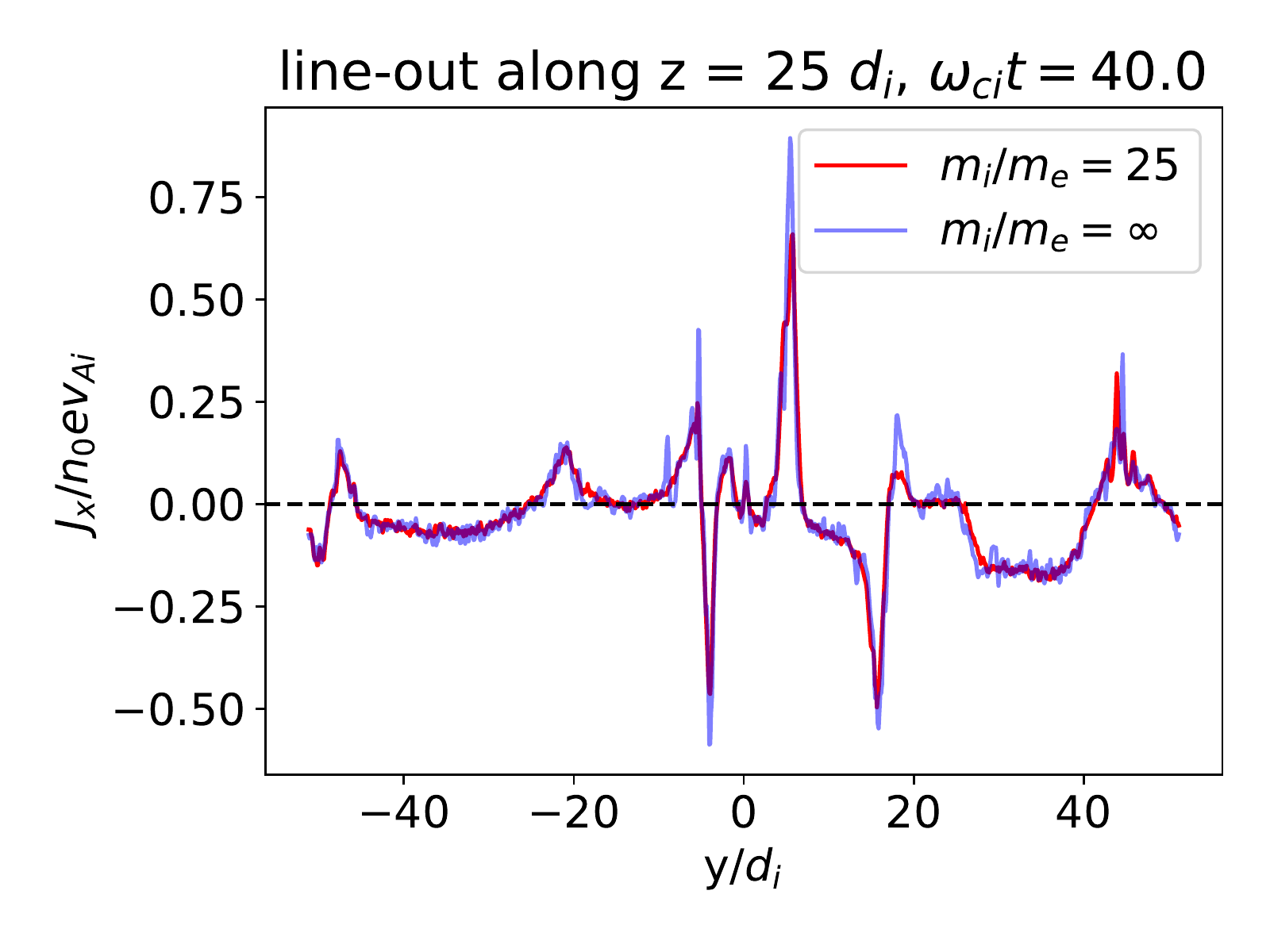}
	\end{minipage}
	\begin{minipage}[b]{0.49\textwidth}
		\includegraphics[width=0.99\textwidth]{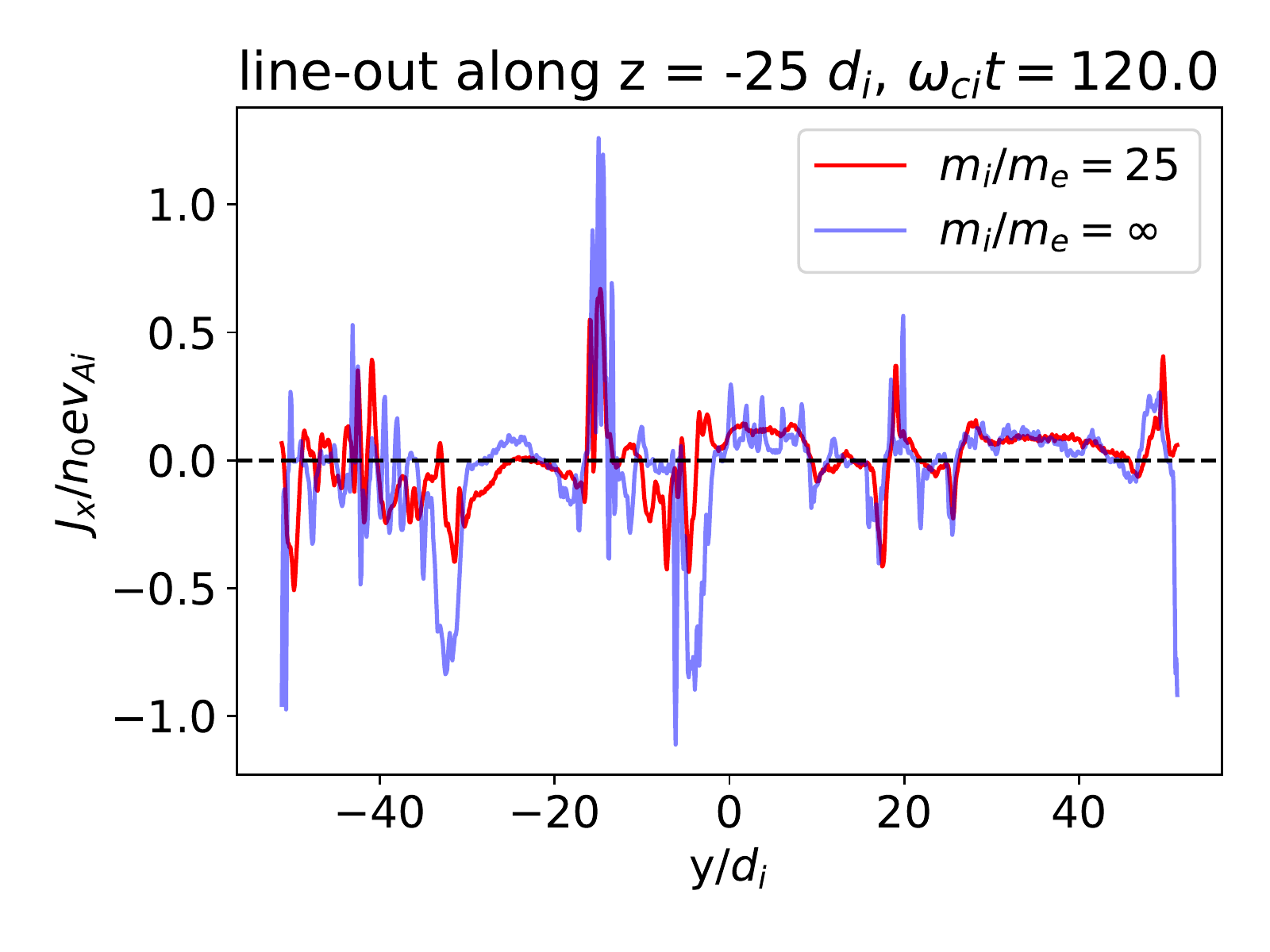}
	\end{minipage}
	\begin{minipage}[b]{0.49\textwidth}
		\includegraphics[width=0.99\textwidth]{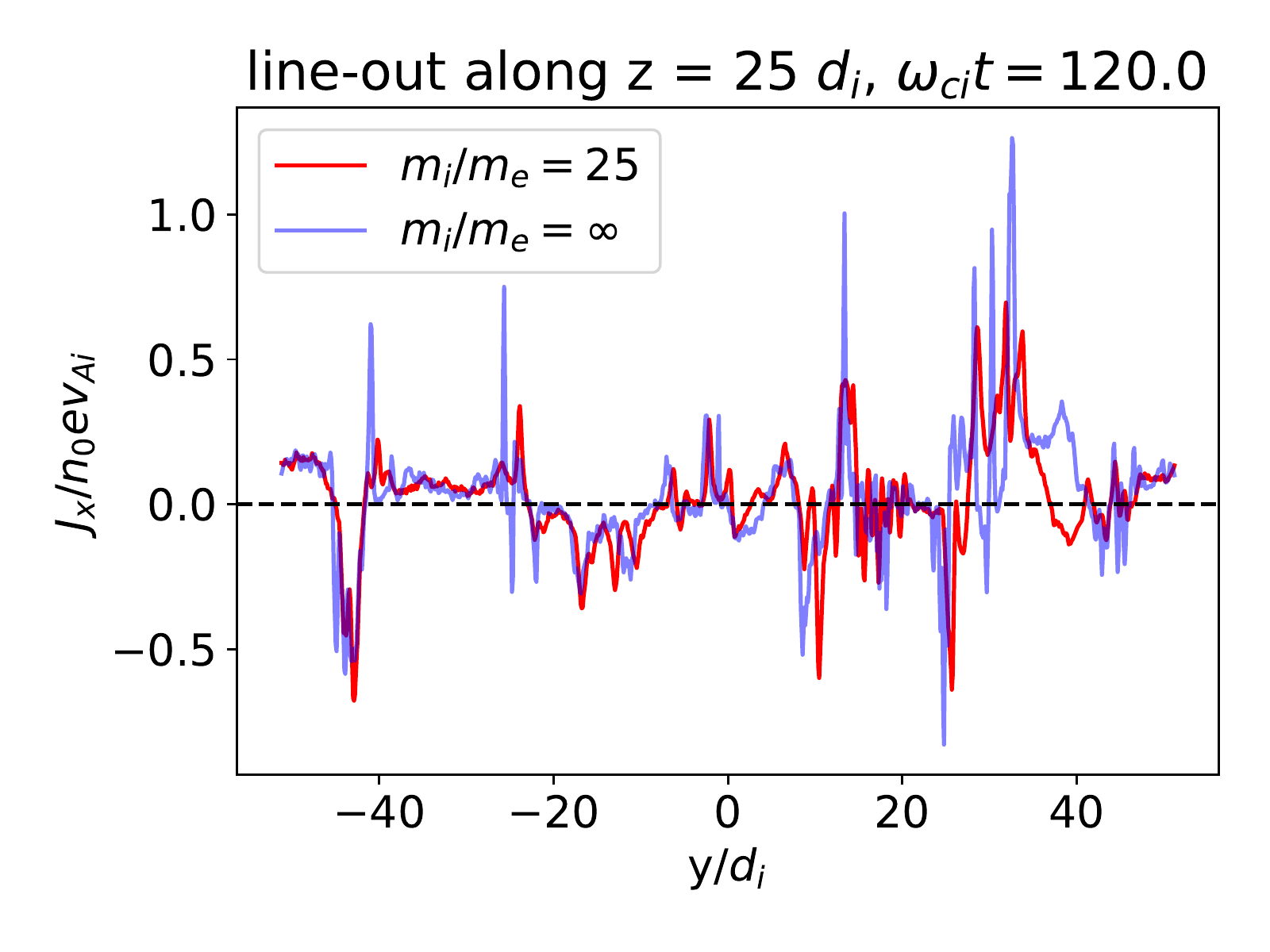}
	\end{minipage}

        \caption{Lineouts along $z/d_i=-25$ (left column) and $z/d_i=25$ (right column) of the out-of-plane current density (shown in Fig.~\ref{fig:jpar_miDme}) for inertial ($m_i/m_e=25$) and inertia-less electrons at $\omega_{ci}t=40$ (top row) and  $\omega_{ci}t=120$ (bottom row). \label{fig:jpar_lineouts}}
\end{figure}

The parallel current density at a given time in Fig. \ref{fig:jpar_miDme} appears to be quite similar for the inertial and inertia-less electrons. They are, however, different at the finer scales as shown by their line-outs along the lines $z/d_i=$-25 and 25 in Fig. \ref{fig:jpar_lineouts}. The current density for the case of the inertia-less electrons is typically noisier and spikier than that for the case of the inertial electrons. At $\omega_{ci}t=40$, the current density for the two cases have more or less the same spatial variation at large scales but different at shorter scales where the current density spikes have higher magnitude for the case of inertia-less electrons. This is due to the role that electron inertia plays at short scales. Current sheets formed in collisionless plasma turbulence tend to thin down to grid scale in the case of inertia-less electrons \cite{jain2021,azizabadi2021}. Finite electron inertia interferes with the thinning process of current structures at scales $k_{\perp}d_e \sim 1$ limiting the CSs thicknesses and thus the amplitude of the current density spikes. At $\omega_{ci}t=120$, the large scale spatial variation of the current density in the two cases can be seen to deviate from each other at selected locations. This could be due to the electron inertia driven instabilities operating in electron scale structures in the case of the inertial electrons.

\subsection{Validity of approximated descriptions of the electron inertia in hybrid-kinetic codes \label{sec:approximations}}

In this section, we check the validity of two different approximations, referred to as A1 and A2 here, used in the numerical implementation of the electron inertial terms in hybrid-kinetic codes. In A1, the non-stationary electron inertia term $(m_e/e)\partial\vec{u}_e/\partial t$ in the generalized Ohm's law, Eq. \eqref{eq:e_ohm}, is neglected when Eq. \eqref{eq:e_ohm} is directly used for the calculation of the electric field ~\cite{Shay1998,Kuznetsova1998,Lipatov2002}. The term is, however, retained to obtain the evolution equation for magnetic field by taking the curl of Eq. \eqref{eq:e_ohm}. This is not entirely self-consistent. We check the effect of neglecting $(m_e/e)\partial\vec{u}_e/\partial t$ in the calculation of electric field from Eq. \eqref{eq:e_ohm} by comparing the results of our simulations with the results of additional simulations carried out by disabling the $(m_e/e)\partial\vec{u}_e/\partial t$ term in Eq. \eqref{eq:e_ohm}.

\begin{figure}[ht!]
	\begin{minipage}[b]{0.49\textwidth}
		\includegraphics[width=0.99\textwidth]{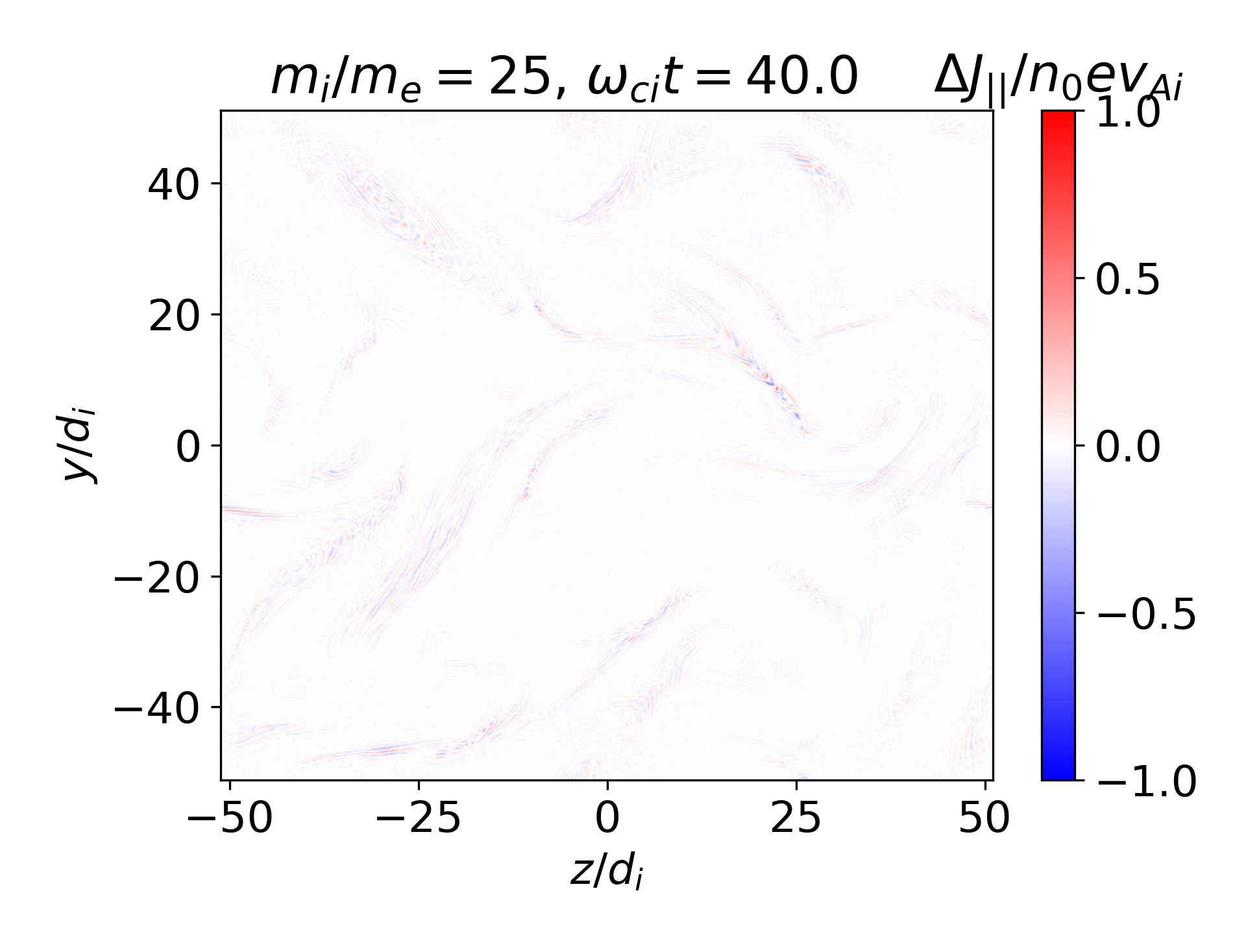}
	\end{minipage}
	\begin{minipage}[b]{0.49\textwidth}
		\includegraphics[width=0.99\textwidth]{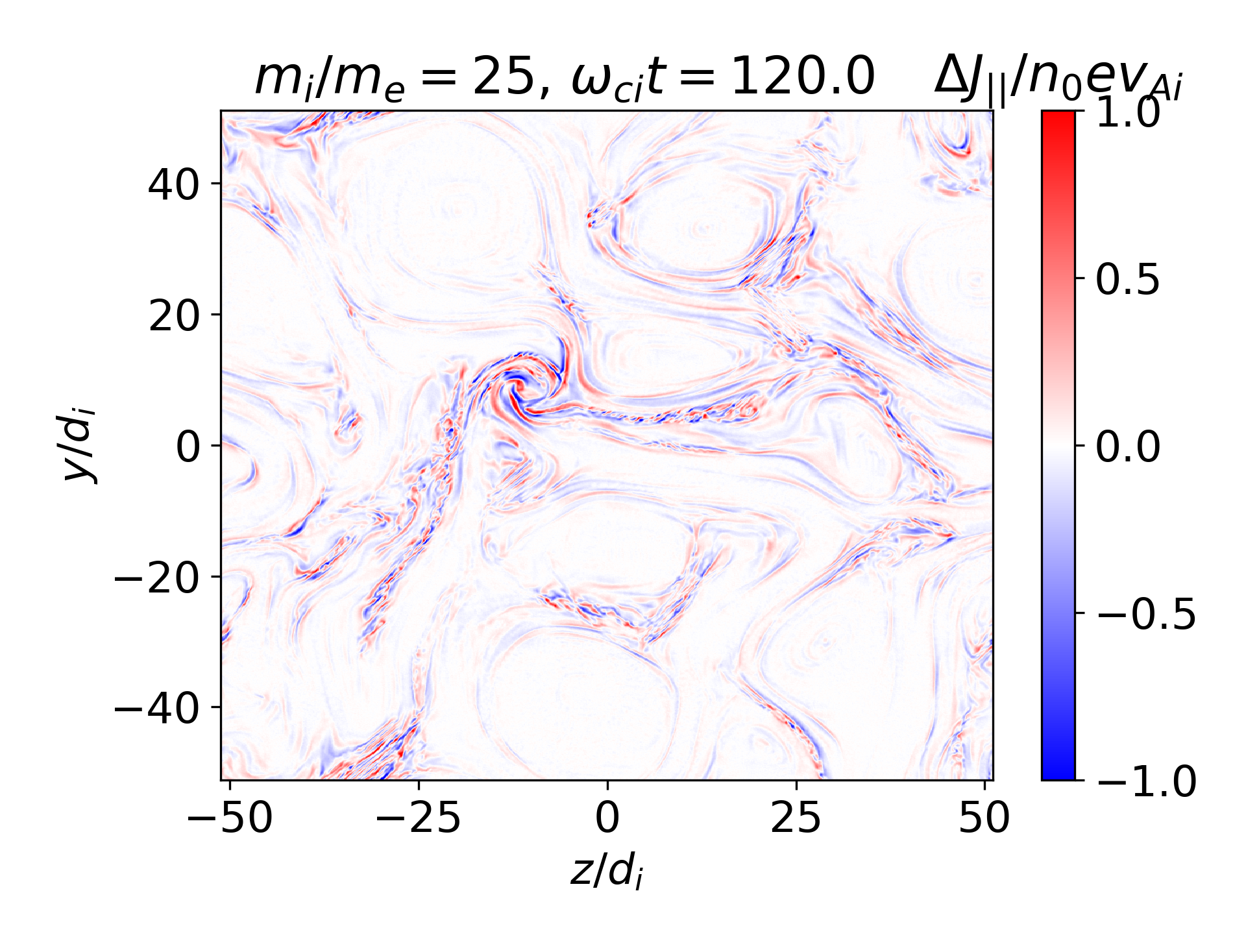}
	\end{minipage}

        \caption{Difference of parallel current densities  $\Delta J_{||}=J_{||}-J_{||}^{\prime}$ at $\omega_{ci}t=40$ (left) and 120 (right). Here $J_{||}^{\prime}$ is the parallel current density obtained from simulations by disabling $(m_e/e)\partial\vec{u}_e/\partial t$ term in Eq. \eqref{eq:e_ohm}. \label{fig:diff_jpar_duedt_disabled}}
\end{figure}

Fig. \ref{fig:diff_jpar_duedt_disabled} shows the 
differences in parallel current densities $\Delta J_{||}=J_{||}-J_{||}^{\prime}$ at $\omega_{ci}t=40$ and 120. Here $J_{||}^{\prime}$ is the parallel current density obtained from simulations additionally carried out by disabling the $(m_e/e)\partial\vec{u}_e/\partial t$ term in Eq. \eqref{eq:e_ohm}. The magnitude of $\Delta J_{||}$ is very small in the whole simulation plane except in electron-scale-thin current sheet like regions where it is a small fraction of $J_{||}$ at $\omega_{ci}t=40$ but becomes a significantly large fraction of $J_{||}$ later at $\omega_{ci}t=120$ (compare Figs. \ref{fig:diff_jpar_duedt_disabled} and \ref{fig:jpar_miDme}). The larger magnitude of $\Delta J_{||}$ at $\omega_{ci}t=120$ is expected as the term $(m_e/e)\partial\vec{u}_e/\partial t$ in Eq. \eqref{eq:e_ohm} is likely to be playing a role only after the electron scale current structure are formed. Thus, the approximation A1 introduces errors at the electron scales in the calculation of the current density. 

The approximation A2 is used by hybrid-kinetic codes which solve an elliptic partial differential equation, instead of Eq. \eqref{eq:e_ohm}, for the electric field.
The elliptic equation for the electric field can be obtained by taking the curl of Faraday's law, using Ampere's law $\nabla \times \vec{B}=\mu_0\vec{J}$ and then substituting for $\partial\vec{J}/\partial t$ which can be obtained from the generalized Ohm's law, 

	\begin{eqnarray}
		\nabla\times\nabla\times\vec{E}&+&\frac{1}{d_e^2}\left(1+\frac{m_e}{m_i}\right) \vec{E} =-\frac{1}{d_e^2}\left(\vec{u}_e+\frac{m_e}{m_i}\vec{u}_i\right)\times\vec{B}\nonumber\\  &+&\frac{1}{ned_e^2}\left[m_e\nabla.[n\left(\vec{u}_i\vec{u}_i-\vec{u}_e\vec{u}_e\right)]
		+\frac{m_e}{m_i}\nabla.\underbar{P}_i-\nabla p_e\right] \label{eq:elliptic_E}.
	\end{eqnarray}

Here $\underbar{P}_i$ is the ion pressure tensor. This elliptic partial differential equation has the advantage that it does not have explicit time derivative terms as in Eq. \eqref{eq:e_ohm}. 
The solution of Eq. \eqref{eq:elliptic_E} is, however, a bit more involved than the solution of Eq. \eqref{eq:elliptic_b} due to the presence of cross-derivative terms in Eq. \eqref{eq:elliptic_E}. The cross derivative terms arise from the $\nabla(\nabla.\vec{E})$ term on the LHS of Eq. \eqref{eq:elliptic_E} as $\nabla\times\nabla\times\vec{E}=\nabla(\nabla.\vec{E})-\nabla^2\vec{E}$. 
Hybrid-kinetic codes using the approximation A2 neglect the $\nabla(\nabla.\vec{E})$ term to obtain a simpler form of the elliptic PDE for the electric field \cite{Valentini2007}. There is, however, no {\it a priori} reason to justify this. 

We check the validity of neglecting the $\nabla(\nabla.\vec{E})$ term in Eq. \eqref{eq:elliptic_E}. Note that our code does not solve Eq. \eqref{eq:elliptic_E} to calculate the electric field. It is, therefore, not possible to directly disable the $\nabla(\nabla.\vec{E})$ term in our simulations and compare the simulation results with the term disabled and enabled, as was done to check the 
approximation A1. We, instead, compare the quantities $\nabla(\nabla.\vec{E})$ and $\nabla^2 \vec{E}$ calculated from the electric field data obtained by solving Eq. \eqref{eq:e_ohm} in our simulations to find their relative importance. 
\begin{figure}[ht!]
	\begin{minipage}[b]{0.49\textwidth}
		\includegraphics[width=0.99\textwidth]{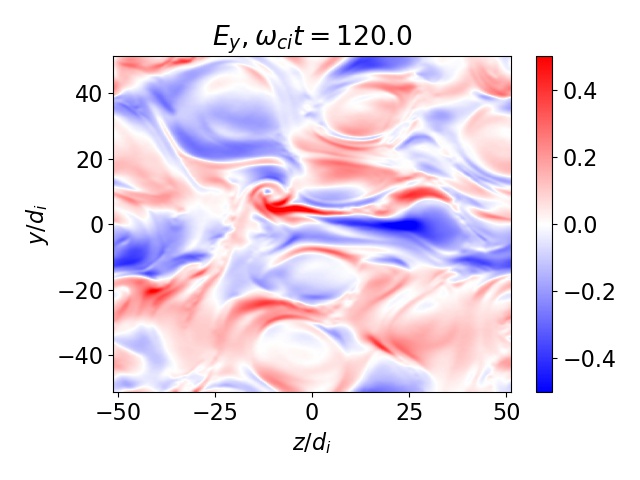}
	\end{minipage}
	\begin{minipage}[b]{0.49\textwidth}
		\includegraphics[width=0.99\textwidth]{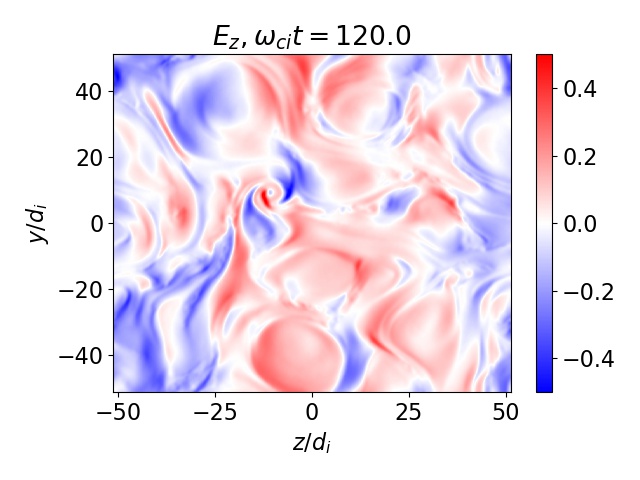}
	\end{minipage}

        \caption{Gaussian-filtered in-plane components $E_y/v_{Ai}B_0$ and $E_z/v_{Ai}B_0$ of the electric field for $m_i/m_e=25$ at $\omega_{ci}t=120$.  \label{fig:ey_ez}}
\end{figure}

Calculations of $\nabla(\nabla.\vec{E})$ and $\nabla^2\vec{E}$ require finding spatial derivatives of the electric field which as obtained from the simulations is noisy, making derivatives noisier. For this reason we smooth the electric field obtained from the simulations by applying to it a Gaussian filter with a standard deviation of four times the grid spacing. Fig. \ref{fig:ey_ez} shows the filtered components of the electric field at $\omega_{ci}t=120$.

\begin{figure}[ht!]
	\begin{minipage}[b]{0.49\textwidth}
		\includegraphics[width=0.99\textwidth]{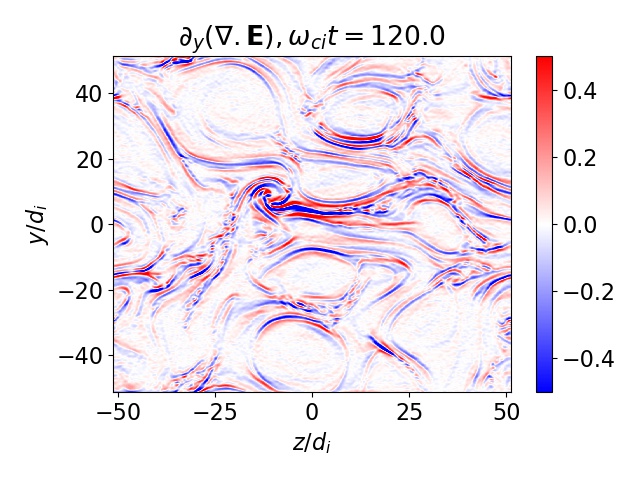}
	\end{minipage}
	\begin{minipage}[b]{0.49\textwidth}
		\includegraphics[width=0.99\textwidth]{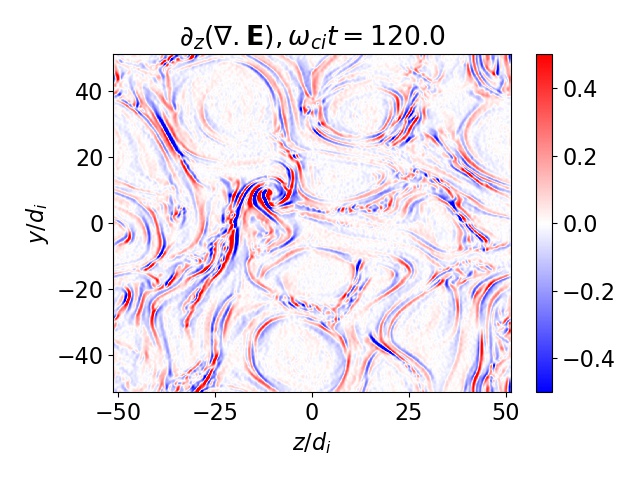}
	\end{minipage}
	\begin{minipage}[b]{0.49\textwidth}
		\includegraphics[width=0.99\textwidth]{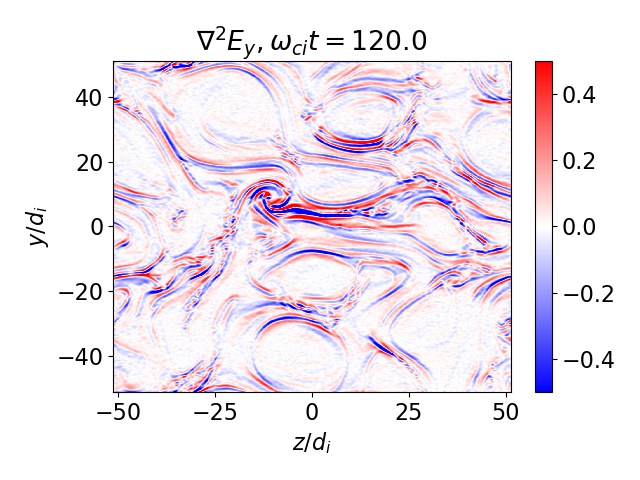}
	\end{minipage}
	\begin{minipage}[b]{0.49\textwidth}
		\includegraphics[width=0.99\textwidth]{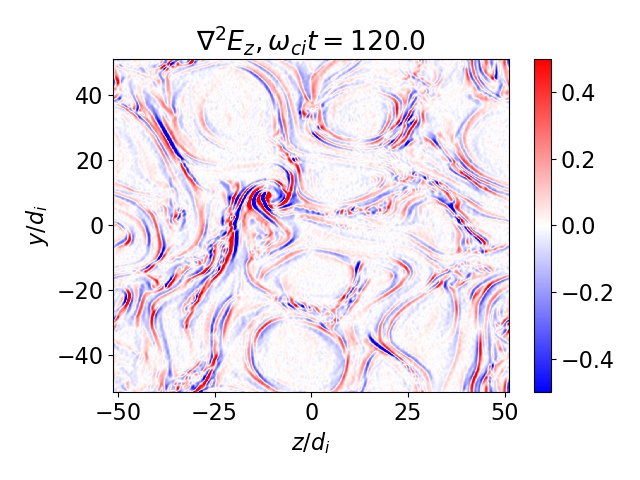}
	\end{minipage}

        \caption{In-plane y- and z-components of $\nabla(\nabla.\vec{E})$ (top row) and $\nabla^2\vec{E}$ (bottom row) obtained by differentiating the Gaussian-filtered electric field at $\omega_{ci}t=120$. \label{fig:electric_field_eq_terms}}
\end{figure}

\begin{figure}
	\begin{minipage}[b]{0.49\textwidth}
		\includegraphics[width=0.99\textwidth]{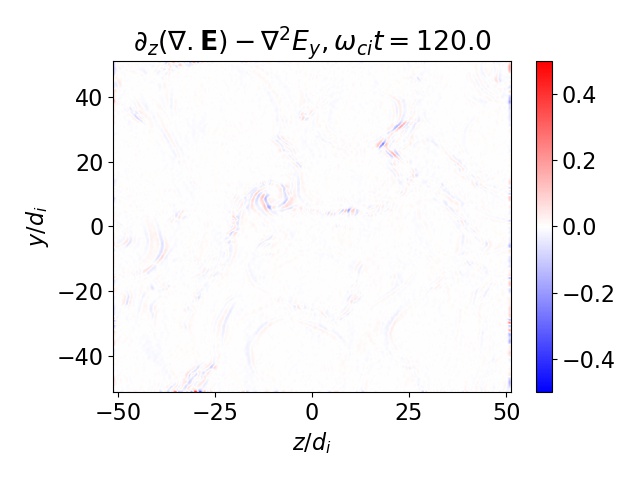}
	\end{minipage}
	\begin{minipage}[b]{0.49\textwidth}
		\includegraphics[width=0.99\textwidth]{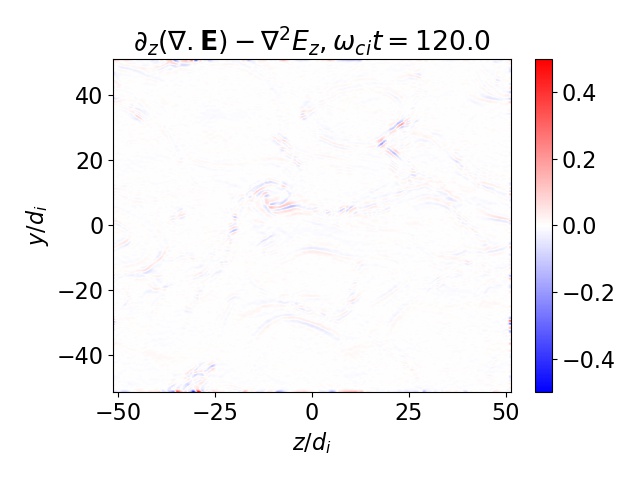}
	\end{minipage}

        \caption{Difference of y-components (left) and z-components (right) of $\nabla(\nabla.\vec{E})$ and $\nabla^2\vec{E}$  at $\omega_{ci}t=120$. \label{fig:diff_electric_field_eq_terms}
        }
\end{figure}

Fig. \ref{fig:electric_field_eq_terms} shows the components of $\nabla(\nabla.\vec{E})$ and $\nabla^2\vec{E}$ at $\omega_{ci}t=120$ calculated from the filtered electric field data. The corresponding components of the two quantities are almost cancelling each other. The differences of the corresponding components, shown in Fig. \ref{fig:diff_electric_field_eq_terms}, is almost negligible everywhere in the y-z plane and only feebly noticeable in some electron scale structures. Therefore, the two quantities are almost equal at all scales and neglecting $\nabla(\nabla.\vec{E})$ while keeping at the same time $\nabla^2\vec{E}$ in Eq. \eqref{eq:elliptic_E} is not consistent.

\begin{figure}[ht!]
	\begin{minipage}[b]{0.49\textwidth}
		\includegraphics[width=0.99\textwidth]{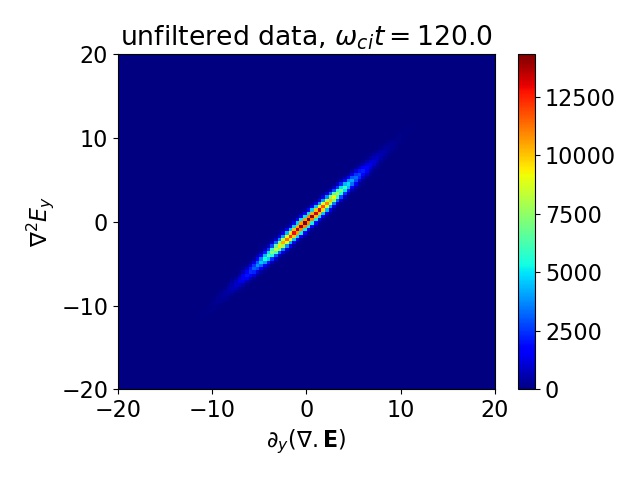}
	\end{minipage}
	\begin{minipage}[b]{0.49\textwidth}
		\includegraphics[width=0.99\textwidth]{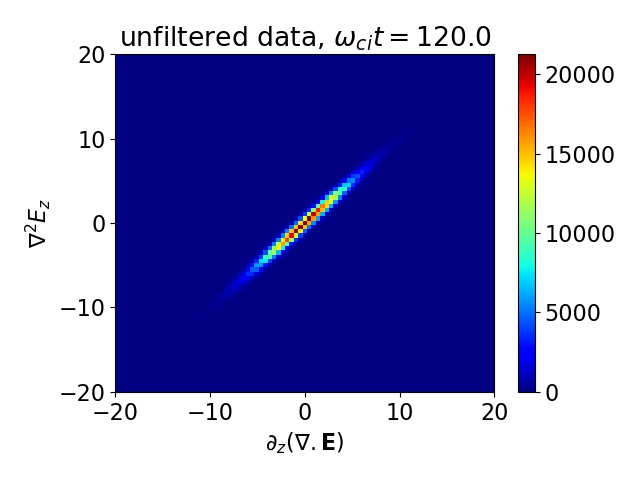}
	\end{minipage}

        \caption{Two dimensional histogram of the corresponding components of  $\nabla(\nabla.\vec{E})$ and $\nabla^2\vec{E}$ obtained using unfiltered data at $\omega_{ci}t=120$. \label{fig:hist2d_electric_field_eq_terms}}
\end{figure}

In order to confirm that the filtering procedure is not contributing to the cancellation of the two quantities, we show in Fig. \ref{fig:hist2d_electric_field_eq_terms} two-dimensional histograms of the corresponding components of the two quantities without applying any filter. It is clear from this figure that they have very good linear correlation, thereby, cancelling each other in Eq. \eqref{eq:elliptic_E}.

Due to the almost perfect cancellation of the two terms $\nabla(\nabla.\vec{E})$ and $\nabla^2\vec{E}$, it is tempting to neglect the term $\nabla\times\nabla\times\vec{E}=\nabla(\nabla.\vec{E})-\nabla^2\vec{E}$ in Eq. \eqref{eq:elliptic_E}. Such a neglect would reduce Eq. \eqref{eq:elliptic_E} to an algebraic equation for the calculation of the electric field, greatly simplifying the numerical algorithm and improving the parallel efficiency of a hybrid code. In order to estimate the impact of neglecting $\nabla\times\nabla\times\vec{E}$ in Eq. \eqref{eq:elliptic_E} using our simulations which solve Eq. \eqref{eq:e_ohm} instead of Eq. \eqref{eq:elliptic_E}, we write $\nabla\times\nabla\times\vec{E}=-\mu_0e\partial(n\vec{u}_i-n\vec{u}_e)/\partial t$ using Faraday's and Ampere's law and approximate it as $\nabla\times\nabla\times\vec{E}=\mu_0en\partial \vec{u}_e / \partial t$ by neglecting the time derivative of the ion bulk velocity and plasma density at electron scales. Therefore, neglecting the $\nabla\times\nabla\times\vec{E}$ term in Eq. \eqref{eq:elliptic_E} (in comparison to the second term on the LHS of Eq. \eqref{eq:elliptic_E}) is equivalent to neglecting $(m_e/e)\partial \vec{u}_e/\partial t$ term on the RHS of Eq. \eqref{eq:e_ohm} (in comparison to $\vec{E}$ on the LHS of Eq. \eqref{eq:e_ohm}). We have already seen in Fig. \ref{fig:diff_jpar_duedt_disabled} that neglecting $(m_e/e)\partial \vec{u}_e/\partial t$ in Eq. \eqref{eq:e_ohm} impacts the results on electron scales. It is, therefore, not appropriate to neglect the $\nabla\times\nabla\times\vec{E}$ term while calculating the electric field from Eq. \eqref{eq:elliptic_E}.

\section{Summary}
\label{sec:conclusion}

We employed our recently updated and parallelized three-dimensional PIC-hybrid code 
CHIEF, which avoids electron inertia related approximations commonly used in hybrid-kinetic codes, to explore the importance of the accurately considered electron inertia.
For this sake we carried out simulations of collisionless plasma turbulence in a 
quasi-two dimensional setup. 
We checked the validity of two approximations made by previously developed hybrid-kinetic codes to implement electron inertial effects. These two approximations, here referred to as A1 and A2, are made in the calculation of electric field by two different algorithms, one of which obtains the electric field directly from the generalized Ohm's law (Eq. \eqref{eq:e_ohm}) and the other via the solution of the elliptic PDE (Eq. \eqref{eq:elliptic_E}).
In the approximation A1 the non-stationary electron inertial term $(m_e/e)\partial \vec{u}_e/\partial t$ in Eq. \eqref{eq:e_ohm} 
is neglected. 
In the approximation A2, $\nabla(\nabla.\vec{E})$ in the expression of $\nabla\times\nabla\times\vec{E}=\nabla(\nabla.\vec{E})-\nabla^2 \vec{E}$ is neglected to replace 
$\nabla\times\nabla\times\vec{E}$ by $-\nabla^2\vec{E}$ 
in Eq. \eqref{eq:elliptic_E}. 
We determined the errors which in approximation A1 are introduced at electron scales. 
On the other hand, we found that approximation A2 is invalid from ion to electron 
scales. Hence, results of hybrid-kinetic simulations of collisionless plasma turbulence 
using either of the two approximations (A1 and A2) need to be interpreted with caution.

A question arises naturally: What is the physical nature of the errors introduced by the approximations 
A1 and A2 into the description of collisionless plasma turbulence? 
Hybrid-kinetic simulations with the approximation A1 may miss the essential physics of collisionless dissipation which occurs mainly in and around electron scale 
current sheets. 
In fact, hybrid-kinetic simulations with an inertia-less electron fluid 
have already shown that ion scale current sheets forming in collisionless turbulent
plasmas develop strong electron shear flows~\cite{jain2021}. 
Our hybrid-kinetic simulations with inertial electron fluid now confirm 
the development of electron shear flows at electron scale current sheets 
(results not shown here). 
The electron shear flow at CSs cause sheet tearing, allow magnetic 
reconnection, and non-tearing plasma instabilities.
For them the inertial term $(m_e/e)\partial \vec{u}_e/\partial t$ in 
Eq. \eqref{eq:e_ohm} plays a crucial role~\cite{jain2015}. 
Two-and-a-half-dimensional fully kinetic PIC code simulations of collisionless 
magnetic reconnection have shown that the inertial term
$(m_e/e)\partial \vec{u}_e/\partial t$ can contribute to the reconnection electric field in electron scale current sheets~\cite{hesse1998}.
The question to what degree avoiding approximation A1 affects collisionless 
dissipation in CSs is now under investigations and will be presented 
separately. 
Since the CHIEF algorithm does not directly solve Eq. \eqref{eq:elliptic_E}, 
studies of the physical impact of approximation A2 cannot be carried out using CHIEF code alone. 
A comparison of the results obtained by CHIEF and those obtained by other hybrid-kinetic codes 
using approximation A2 needs to be carried out for this purpose.

To the best of our knowledge, CHIEF is the first PIC-hybrid code with 
an accurately described inertial electron fluid to simulate collisionless 
plasma turbulence. The computational effort required for CHIEF-code
simulation runs is comparable with that of other hybrid-kinetic codes~\cite{hu2020} (cf. Appendix \ref{sec:performance}) ,
but CHIEF comes with the advantage of the accurate consideration of
the electron inertia. 
Large-scale CHIEF-code simulations in three dimensions and their performance analysis will be presented in a forthcoming 
publication.

\appendix
\section{Code performance and parallel scalability 
\label{sec:performance}}

In order to assess the parallel scalability of our new code a number of benchmark simulations were
performed with CHIEF, using the setup of the model described in Sec.~\ref{sec:setup} and employing different
sizes of the quasi-two dimensional numerical grid ($4\times2048\times 2048$ and $4\times4096\times 4096$). The three-dimensional grid is partitioned across MPI processes using a two-dimensional domain decomposition of the last two dimensions into ``pencils'' with a height of 4 grid cells each.
The number of particles per grid cell was set to 200 and 500.

All calculations were performed on the HPC system \emph{Cobra} at the Max Planck Computing and Data Facility, the compute nodes of which consist of two Intel Xeon Gold 6148 (``Skylake'' architecture) processors with 40 cores per node, operated at the nominal frequency of 2.4 GHz (Turbo mode disabled). The nodes are connected through a 100 Gb/s OmniPath interconnect with a non-blocking fat-tree topology. We used Intel 19.1.3 C++ and Fortran compilers and Intel MPI 2019.7, Intel MKL 2020.4, Blitz++ 1.0.1, and HYPRE 2.22 libraries. Out of several alternatives offered by HYPRE, the hybrid solver (BiCGSTAB with a single step of BoomerAMG preconditioning) was found to provide the best performance for our setup.

\begin{figure}[ht!]
	\begin{minipage}[b]{0.49\textwidth}
		\includegraphics[trim=8cm 0cm 2cm 4cm,width=0.99\textwidth]{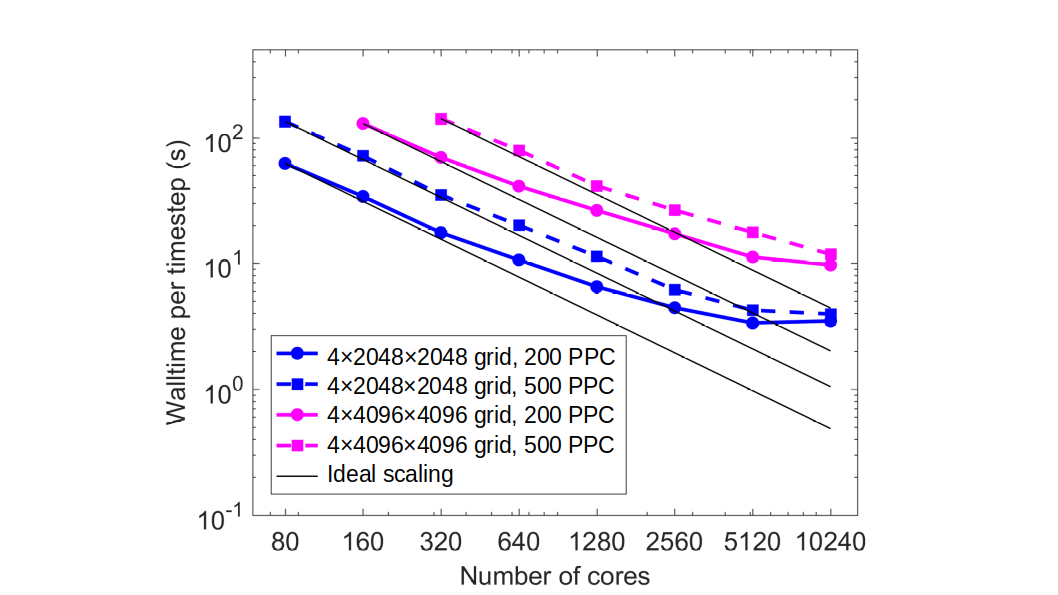}
	\end{minipage}
	\begin{minipage}[b]{0.49\textwidth}
		\includegraphics[width=0.99\textwidth]{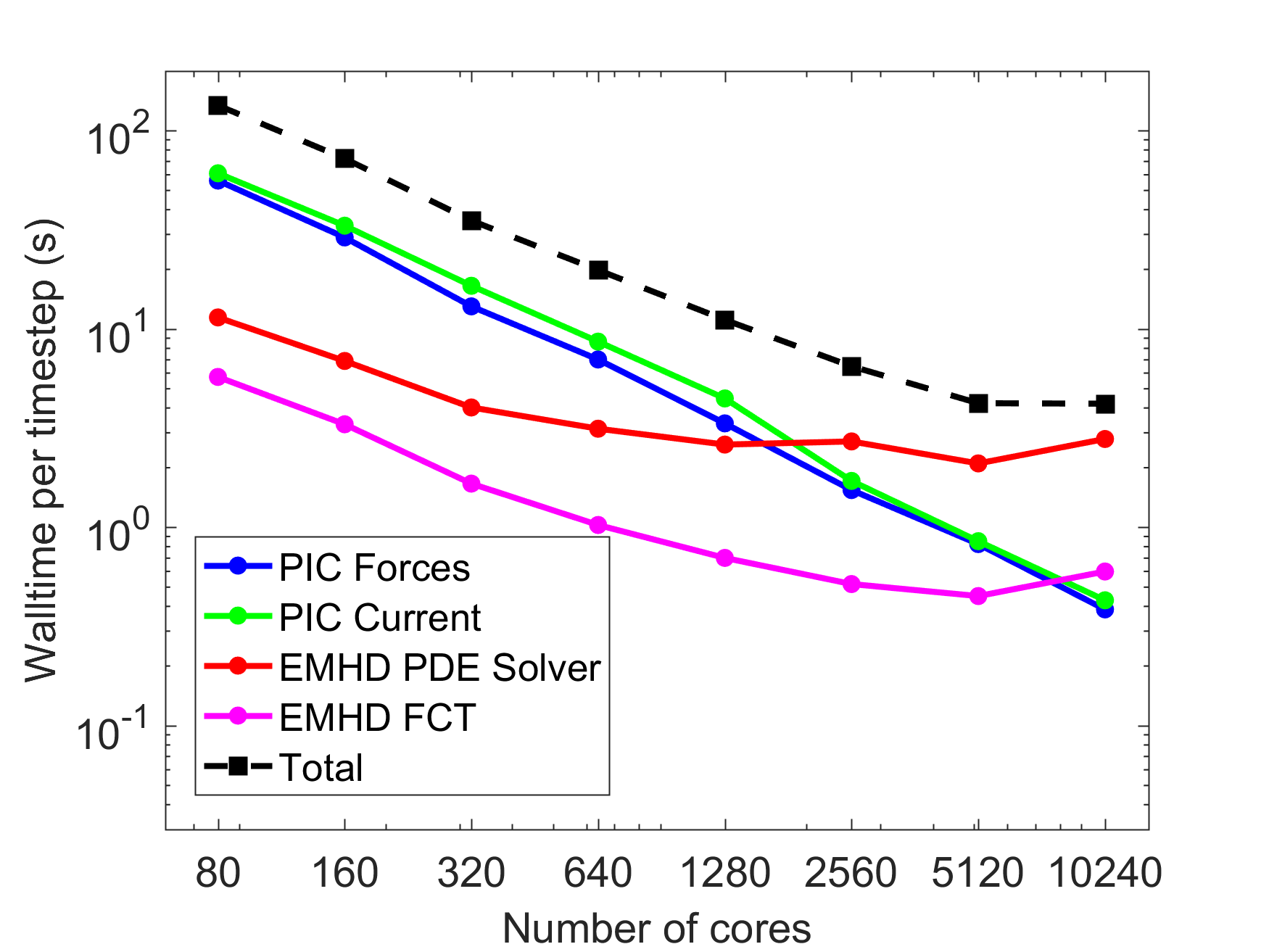}
	\end{minipage}
\caption{Strong scaling results (run time as a function of compute cores) for the CHIEF code using up to 256 nodes (left panel), and a breakdown into its main algorithmic parts for the $4\times 2048\times 2048$ grid with 500 particles per cell (right panel) 
\label{fig:strongscaling}}
\end{figure}

The left panel of Fig.~\ref{fig:strongscaling} shows the run time per time step (average of the first three time steps in 4 different runs) as a function of the number of computing cores (resp. MPI tasks) for two different grid sizes (coloured symbols) up to 10240 cores (256 nodes) with 200 and 500 particles per cell illustrated with solid and dashed-lines, respectively. The black lines indicate ideal strong scaling, taking the computation time on the minimum number of nodes with 768 GB RAM which could run the calculations as the reference. For the larger grid size ($4\times 4096\times 4096$) with 500 particles per cell the code maintains good strong scalability up to roughly 2560 cores, and is able to at least decrease the run time down to ca. 12 s per time step when using 10240 cores. As expected, deviation from close-to-ideal scaling occurs at smaller core counts for smaller grids or smaller number of particles per cell. 

For the grid size of $4\times 2048\times 2048$, the right panel of Fig.~\ref{fig:strongscaling} shows a breakdown of the strong scaling behaviour into the main algorithmic parts (detailed in \citet{Munoz2018}) of CHIEF. When running on a single node, the overall performance of the code is mainly limited by the available memory bandwidth. When increasing the number of nodes, while PIC parts scale virtually perfectly up to 2560 cores (and beyond), the overall strong scalability gets increasingly limited by the less scalable PDE solver in the EMHD part of the code. A detailed analysis has shown that the HYPRE solver is plagued by a significant intrinsic load imbalance beyond a few hundred cores and thus is identified as the main scaling bottleneck for the given setup. 

For 2560 compute cores, the CHIEF code takes approximately 6.5 seconds (or 4.6 core-hours) per time step for the numerical grid of size 4$\times$2048$\times$2048 and 500 particles per cell. This gives a total computational cost of 0.92 million core-hours for a simulation resolving electron time scales by a time step of 0.25 $\omega_{ce}^{-1}$ (0.0025 $\omega_{ci}^{-1}$ for $m_i/m_e=100$)
running for 500 $\omega_{ci}^{-1}$. By comparison, the largest hybrid simulations (known to the authors) of collisionless plasma turbulence carried out using Vlasov-hybrid code covering this scale range for ion to electron mass ratio $m_i/m_e=100$ and a 2048$^2$ numerical grid takes 2 million core-hours \cite{hu2020} (see table 1 in the cited reference). This suggests that our CHIEF code requires a similar computational effort for two dimensional simulations as other hybrid-kinetic codes but CHIEF comes with the additional advantage of the accurate numerical implementation of electron inertial terms. 

\section{Discretization of a 3-D elliptic partial differential equation 
\label{sec:discretizaion_elliptic}}
Most general form of a second order elliptic partial differential equation without any cross-derivative term can be written as,
\begin{eqnarray}
  c_{xx} \frac{\partial^2 u}{\partial x^2} + c_{yy}
  \frac{\partial^2 u}{\partial y^2} + c_{zz} \frac{\partial^2
  u}{\partial z^2} + c_x \frac{\partial u}{\partial x} + c_y \frac{\partial
  u}{\partial y} + c_z \frac{\partial u}{\partial z} + c_e u & = & r 
  \label{eq:elliptic-pde-general-form}
\end{eqnarray}
Here $u$ is the unknown. The coefficients, $c_{xx}$,
$c_{yy}$, $c_{zz}$, $c_x$, $c_y$, $c_z$, $c_e$ and the right
hand side $r$ are all known functions of ($x$, $y$, $z$). 
In the code CHIEF, we solve three elliptic equations, one for each component of the magnetic field (Eq. \eqref{eq:elliptic_b}). In all the three equations $c_{xx} = c_{yy}
= c_{zz} = 1$ and $c_e = - n_e$. The coefficients $c_x$, $c_y$
and $c_z$ are different for the three equations. We, however, discretize the general form in Eq. (\ref{eq:elliptic-pde-general-form}) for the sake of completeness and write the discretized equations in the matrix form suitable for their numerical solution. 

We discretize Eq. \eqref{eq:elliptic-pde-general-form} on a numerical grid consisting of points represented by indices ($i,j,k$) with $i = 1,2, \ldots  M$, $j = 1, 2, \ldots N$ and $k = 1, 2, \ldots  P$. Here $M$, $N$ and $P$ are the number of grid points along $x$, $y$ and $z$ directions, respectively. The two boundary points in each of the $x$, $y$, and $z$ directions correspond to $i=(1, M)$, $j=(1, N)$ and $k=(1, P)$.

Using second order central difference formula for the partial derivatives, Eq. \eqref{eq:elliptic-pde-general-form} can be discretized at a point ($i,j,k$) to yield the following algebraic equation.
\begin{eqnarray}
  \gamma^-_{i, j, k} u_{i, j, k - 1} + \beta^-_{i, j, k} u_{i, j - 1, k} +
  \alpha^-_{i, j, k} u_{i - 1, j, k} + \lambda_{i, j, k} u_{i, j, k} && \nonumber\\
  +\alpha^+_{i, j, k} u_{i + 1, j, k} + \beta^+_{i, j, k} u_{i, j + 1, k} +
  \gamma^+_{i, j, k} u_{i, j, k + 1} &=  &r_{i, j, k} 
  \label{eq:discretized-3d-eq}
\end{eqnarray}
where,
\begin{eqnarray}
  \alpha^{\pm}_{i, j, k} & = & \frac{c_{xx}^{(i, j, k)}}{\Delta_x^2}
  \pm \frac{c_x^{(i, j, k)}}{2 \Delta_x},  \label{alpha}\\
  \beta^{\pm}_{i, j, k} & = & \frac{c_{yy}^{(i, j, k)}}{\Delta_y^2} \pm
  \frac{c_y^{(i, j, k)}}{2 \Delta_y},  \label{beta}\\
  \gamma^{\pm}_{i, j, k} & = & \frac{c_{zz}^{(i, j, k)}}{\Delta_z^2}
  \pm \frac{c_z^{(i, j, k)}}{2 \Delta_z},  \label{gamma}\\
  \lambda_{i, j, k} & = & - \left[ \frac{2 c_{xx}^{(i, j,
  k)}}{\Delta_x^2} + \frac{2 c_{yy}^{(i, j, k)}}{\Delta_y^2} + \frac{2
  c_{zz}^{(i, j, k)}}{\Delta_z^2} - c_e^{(i, j, k)} \right], 
  \label{lambda}
\end{eqnarray}

Here $\Delta_x$, $\Delta_y$ and $\Delta_z$ are the grid spacing in the $x$, $y$ and $z$ directions, respectively. Varying indices over their full range ($i = 1, \ldots M ; j = 1, \ldots N ; k = 1, \ldots P$), Eq. \eqref{eq:discretized-3d-eq} gives $N_t = M N P$ number of equations 
containing terms with $u$ at ghost points (points outside the boundaries)
arising from the index values $i =$ 1 and $M$, $j =$ 1 and $N$, $k =$
1 and $P$, thus making number of unknowns more than the number of
equations. Note that only one of the three indices ($i^*$, $j^*$ and $k^*$) in $u_{i^*,j^*,k^*}$ will be a ghost index (from
the set [$0$, $M + 1$, $N + 1$, $P + 1$]). 
The values of $u$ at ghost points in terms of the internal points have to be
obtained from boundary conditions. We apply periodic boundary conditions in all three directions to obtain ghost values in terms of internal values. 
\begin{eqnarray}
  u_{0, j, k}  =   u_{M-1, j, k} , & &
  u_{M + 1, j, k}  =  u_{1, j, k}   \label{eq:x-ghost-values}\\
  u_{i, 0, k}  = u_{i, N-1, k}, & & 
  u_{i, N + 1, k}  =   u_{i, 1, k}  \label{eq:y-ghost-values}\\
  u_{i, j, 0}  =  u_{i, j, P-1}, && 
  u_{i, j, P + 1}  =  u_{i, j, 1}  \label{eq:z-ghost-values}
\end{eqnarray}
Equations \eqref{eq:x-ghost-values}, \eqref{eq:y-ghost-values} and \eqref{eq:z-ghost-values} are obtained from the application of periodic boundaries in the $x$, $y$ and $z$ directions, respectively. The system of equations, Eq. \eqref{eq:discretized-3d-eq}, can now be written in matrix form. 
\begin{eqnarray}
\overline{\overline{\overline{\lambda}}}\,\,\, \overline{\overline{\overline{u}}} &=&\overline{\overline{\overline{r}}}.
\end{eqnarray}
where $\overline{\overline{\overline{\lambda}}}$, $\overline{\overline{\overline{u}}}$ and $\overline{\overline{\overline{r}}}$ are block matrices of sizes $P\times P$, $P\times 1$ and $P\times 1$, respectively. The number of bars over a variable representing a matrix shows the level of nesting of matrices. A single bar denotes a matrix, double bars a matrix of matrices and triple bars a matrix of matrices of matrices. These block matrices and their sub-matrices are the following. 
\\

$\overline{\overline{\overline{u}}} = \left(\begin{array}{c}
  \overline{\overline{u}}_1\\
  \overline{\overline{u}}_2\\
  \vdots\\
  \vdots\\
  \overline{\overline{u}}_{P - 1}\\
  \overline{\overline{u}}_P
\end{array}\right)$, \,\,\,\,\,
  $\overline{\overline{u}}_k  =  \left(\begin{array}{c}
    \overline{u}_{1, k}\\
    \overline{u}_{2, k}\\
    \vdots\\
    \vdots\\
    \overline{u}_{N - 1, k}\\
    \overline{u}_{N, k}
  \end{array}\right)$, \,\,\,\,\,
$\overline{u}_{j, k} =  \left(\begin{array}{c}
    u_{1, j, k}\\
    u_{2, j, k}\\
    \vdots\\
    \vdots\\
    u_{M - 1, j, k}\\
    u_{M, j, k}
  \end{array}\right)$,\,\,
\\

$\overline{\overline{\overline{r}}} = \left(\begin{array}{c}
  \overline{\overline{r}}_1\\
  \overline{\overline{r}}_2\\
  \vdots\\
  \vdots\\
  \overline{\overline{r}}_{P - 1}\\
  \overline{\overline{r}}_P
\end{array}\right)$, \,\,\,\,\,
$\overline{\overline{r}}_k  =  \left(\begin{array}{c}
    \overline{r}_{1, k}\\
    \overline{r}_{2, k}\\
    \vdots\\
    \vdots\\
    \overline{r}_{N - 1, k}\\
    \overline{r}_{N, k}
  \end{array}\right)$,\,\,\, \,\,
  $\overline{r}_{j, k}  =  \left(\begin{array}{c}
    r_{1, j, k}\\
    r_{2, j, k}\\
    \vdots\\
    \vdots\\
    r_{M - 1, j, k}\\
    r_{M, j, k}
  \end{array}\right)$,\,\,

$\overline{\overline{\overline{\lambda}}}  =  \left(\begin{array}{cccccccc}
    \overline{\overline{\lambda}}_1 & \overline{\overline{\gamma}}^+_1 & 0 & 0 & 0 & \ldots & \overline{\overline{\gamma}}_1^- I & 0\\
    \overline{\overline{\gamma}}_2^- & \overline{\overline{\lambda}}_2 &
    \overline{\overline{\gamma}}^+_2 & 0 & 0 & . & 0 & 0\\
    0 & \overline{\overline{\gamma}}_3^- &\overline{\overline{\lambda}}_3 & \overline{\overline{\gamma}}_3^+ & 0 & . & 0 & 0\\
    . & . & \ddots & \ddots & \ddots & . & . & .\\
    . & . & . & \ddots & \ddots & \ddots & . & .\\
    0 & 0 & 0 & . & \overline{\overline{\gamma}}_{P - 2}^- &
    \overline{\overline{\lambda}}_{P - 2} & \overline{\overline{\gamma}}_{P - 2}^+ & 0\\
    0 & 0 & 0 & . & 0 & \overline{\overline{\gamma}}_{P - 1}^- &
    \overline{\overline{\lambda}}_{P - 1} & \overline{\overline{\gamma}}_{P - 1}^+\\
    0 & \overline{\overline{\gamma}}_P^+ I & 0 & \ldots & 0 & 0 &
    \overline{\overline{\gamma}}_P^- & \overline{\overline{\lambda}}_P
  \end{array}\right)$

$\overline{\overline{\lambda}}_k  =  \left(\begin{array}{cccccccc}
    \overline{\lambda}_{1,k} & \overline{\beta}^+_{1, k} &
    0 & 0 & 0 & \ldots & \overline{\beta}_{1,k}^- I & 0\\
    \overline{\beta}_{2, k}^- & \overline{\lambda}_{2, k} & \overline{\beta}^+_{2, k} & 0 & 0 & . & 0 & 0\\
    0 & \overline{\beta}_{3, k}^- & \overline{\lambda}_{3, k} &
    \overline{\beta}_{3, k}^+ & 0 & . & 0 & 0\\
    . & . & \ddots & \ddots & \ddots & . & . & .\\
    . & . & . & \ddots & \ddots & \ddots & . & .\\
    0 & 0 & 0 & . & \overline{\beta}_{N - 2, k}^- &
    \overline{\lambda}_{N - 2, k} & \overline{\beta}_{N - 2,k}^+ & 0\\
    0 & 0 & 0 & . & 0 & \overline{\beta}_{N - 1, k}^- &
    \overline{\lambda}_{N - 1, k} & \overline{\beta}_{N - 1,k}^+\\
    0 & \overline{\beta}_{N, k}^+ I & 0 & \ldots & 0 & 0
    & \overline{\beta}_{N, k}^- & \overline{\lambda}_{N, k}
  \end{array}\right)$,
\vspace{0.2in}\\
$\overline{\lambda}_{j, k}  =  \left(\begin{array}{cccccccc}
    \lambda_{1, j, k} & \alpha^+_{1, j, k} &
    0 & 0 & 0 & \ldots & \alpha^-_{1, j, k} & 0\\
    \alpha_{2, j, k}^- & \lambda_{2, j, k} &
    \alpha^+_{2, j, k} & 0 & 0 & . & 0 & 0\\
    0 & \alpha_{3, j, k}^- & \lambda_{3, j, k} &
    \alpha_{3, j, k}^+ & 0 & . & 0 & 0\\
    . & . & \ddots & \ddots & \ddots & . & . & .\\
    . & . & . & \ddots & \ddots & \ddots & . & .\\
    0 & 0 & 0 & . & \alpha_{M - 2, j, k}^- &
    \lambda_{M - 2, j, k} & \alpha_{M - 2, j, k}^+ & 0\\
    0 & 0 & 0 & . & 0 & \alpha_{M - 1, j, k}^- &
    \lambda_{M - 1, j, k} & \alpha_{M - 1, j, k}^+\\
    0 & \alpha_{M, j, k}^+ & 0 & \ldots
    & 0 & 0 & \alpha_{M, j, k}^- & \lambda_{M, j, k}
  \end{array}\right)$
\vspace{0.2in}\\
$\overline{\overline{\gamma}}_k^{\pm}  = 
  \left(\begin{array}{ccccccc}
    \overline{\gamma}_{1, k}^{\pm} & 0 & 0 & . & . & 0 & 0\\
    0 & \overline{\gamma}_{2, k}^{\pm} & 0 & . & . & 0 & 0\\
    0 & 0 & \overline{\gamma}_{3, k}^{\pm} & . & . & 0 & 0\\
    . & . & . & \ddots & . & . & .\\
    . & . & . & . & \ddots & . & .\\
    0 & 0 & 0 & . & . &\overline{\gamma}^{\pm}_{N - 1, k} & 0\\
    0 & 0 & 0 & . & . & 0 & \overline{\gamma}_{N, k}^{\pm}
  \end{array}\right)$
\vspace{0.2in}\\
$\overline{\gamma}^{\pm}_{j, k}  =  \left(\begin{array}{ccccccc}
    \gamma_{1, j, k}^{\pm} & 0 & 0 & . & . & 0 & 0\\
    0 & \gamma_{2, j, k}^{\pm} & 0 & . & . & 0 & 0\\
    0 & 0 & \gamma_{3, j, k}^{\pm} & . & . & 0 & 0\\
    . & . & . & \ddots & . & . & .\\
    . & . & . & . & \ddots & . & .\\
    0 & 0 & 0 & . & . & \gamma^{\pm}_{M - 1, j, k} & 0\\
    0 & 0 & 0 & . & . & 0 & \gamma_{M, j, k}^{\pm}
  \end{array}\right)$

$\begin{array}{lll}
  \overline{\beta}_{j, k}^{\pm}  =  \left(\begin{array}{ccccccc}
  \beta_{1, j, k}^{\pm} & 0 & 0 & . & . & 0 & 0\\
    0 & \beta_{2, j, k}^{\pm} & 0 & . & . & 0 & 0\\
    0 & 0 & \beta_{3, j, k}^{\pm} & . & . & 0 & 0\\
    . & . & . & \ddots & . & . & .\\
    . & . & . & . & \ddots & . & .\\
    0 & 0 & 0 & . & . & \beta^{\pm}_{M - 1, j, k} & 0\\
    0 & 0 & 0 & . & . & 0 & \beta_{M, j, k}^{\pm}
  \end{array}\right)
\end{array}$

Here $j=1,2,...N$ and $k=1,2,...P$. In the hybrid code CHIEF, we solve system of equations, Eq. \eqref{eq:discretized-3d-eq}, using HYPRE package of scalable linear solvers \cite{falgout2002}.   

\begin{acknowledgments}

The authors acknowledge the support by the German Science Foundation (DFG) 
projects JA 2680-2-1, MU-4255/1-1, BU 777-16-1 and BU 777-17-1. 
We acknowledge the developers of the ACRONYM code 
(Verein zur F\"orderung kinetischer Plasmasimulationen e.V.).

\end{acknowledgments}

\bibliography{2DparallelCHIEF}

\end{document}